\documentclass[journal]{IEEEtran}

\usepackage{cite}
\usepackage{amsmath,amssymb,amsfonts}
\usepackage{graphicx}
\usepackage{textcomp}
\usepackage{xcolor}
\usepackage{color}
\usepackage{array}
\usepackage[caption=false,font=footnotesize]{subfig}
\usepackage[hyphens]{url}
\usepackage{multirow}
\usepackage{subfloat}
\usepackage{comment}
\usepackage{tikz}
\usepackage[dvipsnames]{xcolor}
\usepackage[linesnumbered,ruled,vlined]{algorithm2e}
\usepackage[noend]{algpseudocode}
\usepackage{breqn}
\usepackage{dblfloatfix}
\usepackage{subfig}
\usepackage{bigdelim} 
\usepackage{adjustbox}
\usepackage{booktabs}

\usepackage{listings}
\usepackage{xcolor}

\lstdefinelanguage{HJSON}{
    morestring=[b]",
    morecomment=[l]{//},
    sensitive=true
}

\lstdefinestyle{hjsonstyle}{
    language=HJSON,
    basicstyle=\ttfamily\footnotesize,
    numbers=left,
    numberstyle=\scriptsize,
    numbersep=6pt,
    frame=single,
    breaklines=true,
    breakatwhitespace=false,
    showstringspaces=false,
    tabsize=2,
    columns=fullflexible,
    captionpos=b,
    xleftmargin=1em,
    framexleftmargin=-0.2em
}

\lstdefinestyle{shellstyle}{
    language=bash,
    basicstyle=\ttfamily\footnotesize,
    numbers=left,
    numberstyle=\scriptsize,
    numbersep=6pt,
    frame=single,
    breaklines=true,
    showstringspaces=false,
    captionpos=b
}

\lstdefinestyle{SimOutput}{
    basicstyle=\ttfamily\footnotesize,
    frame=single,
    rulecolor=\color{black},
    backgroundcolor=\color{white},
    numbers=left,
    numberstyle=\scriptsize,
    stepnumber=1,
    numbersep=6pt,
    breaklines=true,
    breakatwhitespace=false,
    columns=fullflexible,
    keepspaces=true,
    showstringspaces=false,
    tabsize=2,
    captionpos=b,
    xleftmargin=1em,
    framexleftmargin=-0.2em
}

\def\BibTeX{{\rm B\kern-.05em{\sc i\kern-.025em b}\kern-.08em
    T\kern-.1667em\lower.7ex\hbox{E}\kern-.125emX}}

\makeatletter
\newcommand\footnoteref[1]{\protected@xdef\@thefnmark{\ref{#1}}\@footnotemark}
\makeatother

\begin{document}

\title{Lessons from the Hardware Hacking Competitions: Verification Techniques, Findings, and Insights
}
\author{\IEEEauthorblockN{Sudipta Paria\IEEEauthorrefmark{1}, Aritra Dasgupta\IEEEauthorrefmark{1}, Raghul Saravanan\IEEEauthorrefmark{2}, Jayanth Thangellamudi\IEEEauthorrefmark{2},\\Sai Manoj P D\IEEEauthorrefmark{2}, and Swarup Bhunia\IEEEauthorrefmark{1}}\\
\IEEEauthorblockA{\textit{\IEEEauthorrefmark{1}Department of Electrical and Computer Engineering,}
\textit{University of Florida, Gainesville, FL, USA}\\
\textit{\IEEEauthorrefmark{2}Department of Electrical and Computer Engineering,}
\textit{George Mason University, Fairfax, VA, USA}
\\
\{sudiptaparia,aritradasgupta\}@ufl.edu, \{rsaravan,jthangel,spudukot\}@gmu.edu, swarup@ece.ufl.edu
}
}

\maketitle


\begin{abstract}
Hardware hacking competitions have emerged as practical platforms for evaluating security weaknesses in complex System-on-Chip (SoC) designs while promoting security-aware verification and tool development. This paper presents a systematic study of SoC security verification through open-box hardware hacking competitions, focusing on practical vulnerability analysis strategies, observed findings, and lessons for security-aware verification. We present a multi-strategy vulnerability analysis methodology, combining simulation-based verification, formal verification, lint analysis, Large Language Model (LLM)-assisted bug detection, and coverage-guided hybrid fuzzing. Representative vulnerability findings are analyzed to illustrate how different techniques expose complementary classes of security flaws, and we derive practical lessons for pre-silicon security verification. Finally, we discuss how competition benchmarks can support the reproducible evaluation of emerging hardware security techniques and guide future security-aware EDA research.
\end{abstract}
\vspace{-1em}
\begin{IEEEkeywords}
Hardware Security, Capture The Flag, System-on-Chip Security, RTL, Security Verification, Hack@DAC, Hack@DATE, Hack@CHES, Vulnerability Detection, CWE.
\end{IEEEkeywords}


\vspace{-1em}
\section{Introduction}
\label{s1_intro}

The security of electronic systems fundamentally depends on the security guarantees provided by their underlying hardware stack. Traditionally, system security has relied on the assumption that the underlying hardware behaves correctly and acts as an immutable root of trust, focusing defensive efforts almost exclusively on software vulnerabilities. However, recent microarchitectural and physical exploits such as Meltdown, Spectre, and  power/fault injection attacks have demonstrated that hardware design flaws can completely undermine software-level isolation mechanisms \cite{lou2021survey,hardfails}. Additionally, access control violations, inadequate privilege checking, incorrect key management, improper debug authentication, etc. may affect the security of the entire system. Vulnerabilities embedded within hardware implementation, either at the register transfer level (RTL) or gate-level netlist, cannot be easily patched post-silicon and remain exploitable after fabrication. Unlike software defects, such hardware flaws may require complex workarounds, firmware restrictions, or hardware redesign \cite{hardfails}. Consequently, the semiconductor industry is aggressively pursuing a shift-left strategy, which aims to identify and remediate hardware security weaknesses early in the pre-silicon RTL design phase.
Despite the critical need for pre-silicon assurance, traditional electronic design automation (EDA) environments lack robust, security-aware verification flows. Standard functional verification pipelines are optimized to validate whether a design performs its intended architectural specifications, but they rarely verify whether the design permits unintended, malicious behaviors. Manual RTL review, simulation-based verification, formal verification, static analysis, information-flow tracking, and hardware fuzzing have been proposed as complementary methodologies for detecting design defects and potential security vulnerabilities during the pre-silicon verification phase. 
However, the efficacy of the current verification methodologies depends on the generation of relevant security properties or assertions, test vectors, or automated tool flows to identify subtle corner-case errors, especially in complex or large-scale designs.

Hardware hacking competitions provide a practical setting for addressing security verification challenges, allowing participants to analyze vulnerable designs under a defined threat model and within a limited period. 
This is effective for detecting previously unknown bugs and identifying specialized verification approaches because it exposes hardware designs to diverse adversarial strategies, revealing vulnerabilities that may not surface under traditional verification methodologies. The \textit{hackthesilicon} initiative \cite{hackthesilicon}, including Hack@DAC, Hack@CHES, and Hack@DATE, co-located with highly regarded academic conferences, was established as open-box hardware hacking competitions. These competitions provide academic and industrial participants with a realistic, open-source SoC benchmark containing deliberately inserted CVE/CWE-inspired vulnerabilities. Unlike traditional software Capture the Flag (CTF) events or physical hardware hacking competitions that treat the target as a closed box, these hackathons provide participants with full visibility into a buggy SoC RTL design, its specification, and its intended threat model. Participants are allowed to utilize manual code reviews, formal property verification, dynamic simulation, and emerging automated tools to identify weaknesses, determine root causes, develop triggers or exploits, and evaluate the security impact. The open-box model supports early analysis of security weaknesses and promotes the detection of vulnerabilities during the design stage. The three major goals are identified as: (1) improving awareness of hardware weaknesses, (2) supporting security-aware design automation, and (3) shifting vulnerability detection toward the RTL design phase \cite{kanuparthi2024hackdac}.

Existing studies \cite{hackdac_2021,helloctf,tan2020benchmarkingfrontierhardwaresecurity,sst_competition} have discussed the organization of hardware security competitions, the construction of challenge designs, and individual methods developed for evaluating respective benchmarks. However, they are generally limited to particular methodologies or threat models and provide limited guidance on systematically analyzing large SoCs, assessing candidate vulnerabilities, and validating identified bugs.
The goal of this work is to formalize systematic methodologies for identifying subtle pre-silicon vulnerabilities, discuss corresponding mitigation strategies, and summarize practical lessons learned. By connecting adversarial analysis with hardware verification practices, this paper highlights how open-box competitions can support the development and evaluation of automated security-verification techniques for stronger hardware assurance. 
The main contributions of this paper include:

\begin{itemize}
    \item We present a practical workflow for analyzing vulnerable SoCs from hardware hacking competitions, covering environment setup, threat modeling, vulnerability discovery and validation, exploit development, and bug analysis.
    \item We present representative security vulnerabilities identified across multiple IPs, including their location, security weaknesses, validation methods, and observed impact.
    \item We summarize technical lessons on verification-method selection, attack-surface prioritization, vulnerability validation, and effective use of complementary security-analysis techniques.
    \item We discuss the broader role of hardware hacking competitions in security education, benchmark development, tool evaluation, and hardware security research.
\end{itemize}

The remainder of this paper is organized as follows. Section \ref{sec:background} provides background on SoC security assurance and the evolution of hardware-security competitions. Section \ref{sec:methodology} presents different security analysis methodologies, including simulation, formal verification, lint analysis, LLM-assisted vulnerability analysis, fuzzing, along with the identified bug classes and distribution. Section \ref{sec:lessons} summarizes the key lessons learned from the competitions, while Section \ref{sec:impact} discusses the broader community impact, open challenges, and future research directions. Section \ref{sec:conclusion} concludes the paper.

\section{Background}
\label{sec:background}

\subsection{Security Assurance for SoCs}

Modern SoC designs integrate a complex mix of third-party IP blocks and in-house cores connected via shared on-chip fabrics. This design paradigm introduces significant security verification challenges. If a single IP block contains an access control flaw or a hidden hardware Trojan, the security boundary of the entire system can be compromised. Hence, security verification holds utmost importance in modern IC design flow. SoC security also depends on the correct implementation of system-wide protection mechanisms. These mechanisms include secure boot, hardware root of trust, debug authentication, memory protection, key management, lifecycle control, secure data erasure, and error or alert handling. Such features usually involve several hardware and firmware components. A local implementation error can therefore create a system-level vulnerability. Conventional functional verification is necessary but not sufficient for establishing security. Functional verification primarily evaluates specified design behavior, whereas security verification additionally considers adversarial inputs, privilege violations, invalid states, and cross-component interactions.

\vspace{-1em}
\subsection{Evolution of Hardware Security Competitions}

Hardware-security competitions provide a controlled setting for studying vulnerabilities in processors, system-on-chip (SoC) designs, firmware, and physical devices. These events combine security analysis with hands-on design and verification tasks. Depending on the competition, participants may analyze a fabricated device, inspect RTL source code, develop an exploit, or propose a mitigation.

\subsubsection{Software CTFs vs. Hardware CTFs}

Capture-the-flag competitions are widely used in software and system security \cite{defcon_ctf,plaidctf}. Participants solve security challenges and obtain a flag or other proof that the challenge has been completed. Common tasks include software exploitation, reverse engineering, cryptography, web security, and digital forensics.
Hardware security competitions \cite{whibox_contest,csaw_esc,csaw_ai_hardware_challenge,hackthesilicon} use a related model but often require different forms of evidence. A hardware weakness may not produce a simple textual flag. Participants may instead need to provide:
the affected RTL module or hardware component;
the security property that is violated;
the sequence required to trigger the weakness;
simulation, formal, or physical evidence;
the resulting security impact; and
a proposed mitigation.
Therefore, some hardware hacking events are closer to structured vulnerability-discovery exercises than conventional flag-based CTFs. The result is usually a technical report, exploit, waveform, counterexample, or proof-of-concept test.
Hardware competitions also require additional hardware-specific skills, including experience in RTL design, computer architecture, firmware, simulation, formal verification, FPGA prototyping, physical attacks, and security analysis. Designers can explain the intended hardware behavior, verification engineers can construct test cases and properties, and security researchers can identify adversarial conditions that may not be considered during normal functional verification.

\begin{table*}[!htbp]
\centering
\caption{Comparison between closed vs. open-box hardware CTFs}
\label{tab:closed_open_ctfs}
\resizebox{\textwidth}{!}{%
\begin{tabular}{|l|l|l|}
\hline
\multicolumn{1}{|c|}{\textbf{Characteristic}} & \multicolumn{1}{c|}{\textbf{Closed-box competition}}                 & \multicolumn{1}{c|}{\textbf{Open-box competition}}                     \\ \hline
Primary target                                & Fabricated chip or physical board                                    & RTL implementation of a processor or SoC                                   \\ \hline
Internal visibility                           & Limited or unavailable                                               & Design hierarchy and RTL are available                                     \\ \hline
Common interfaces                             & Debug ports, communication links, firmware, physical signals         & RTL modules, registers, buses, firmware, specifications                    \\ \hline
Main techniques                               & Probing, fault injection, side-channel analysis, reverse engineering & Simulation, formal verification, static analysis, emulation, manual review \\ \hline
Typical attacker view                         & External or physical attacker                                        & Design-time adversary / security analyst                                              \\ \hline
Main evidence                                 & Device output, extracted data, fault response, side-channel trace    & Waveform, assertion failure, formal counterexample, exploit test           \\ \hline
Primary lifecycle stage                       & Post-silicon                                                         & Pre-silicon                                                                \\ \hline
Main outcome                                  & Demonstrated attack against a device                                 & Root-cause analysis and design-level mitigation                            \\ \hline
\end{tabular}%
}
\vspace{-1em}
\end{table*}

\begin{table*}[!ht]
\centering
\caption{Comparison of open-box hardware security competitions}
\label{tab:comparison}
\resizebox{0.98\textwidth}{!}{%
\begin{tabular}{|c|c|c|c|c|}
\hline
\textbf{Competition} &
\textbf{Target and access model} &
\textbf{Participant task} &
\textbf{Common techniques} &
\textbf{Primary focus} \\
\hline

\begin{tabular}[c]{@{}c@{}}CSAW Logic Locking \\Conquest (LLC)\end{tabular} &
\begin{tabular}[c]{@{}c@{}}Locked or redacted RTL and\\gate-level designs\end{tabular} &
\begin{tabular}[c]{@{}c@{}}Recover keys or functionality\\and evaluate protection methods\end{tabular} &
\begin{tabular}[c]{@{}c@{}}SAT attacks, oracle-less\\attacks, structural analysis,\\machine learning\end{tabular} &
\begin{tabular}[c]{@{}c@{}}Hardware IP protection\\benchmarking\end{tabular} \\
\hline

\begin{tabular}[c]{@{}c@{}}CSAW AI Hardware \\Attack Challenge\end{tabular} &
\begin{tabular}[c]{@{}c@{}}Open-source RTL designs with\\required AI-assisted modification\end{tabular} &
\begin{tabular}[c]{@{}c@{}}Insert and exploit hardware\\Trojans or security weaknesses\end{tabular} &
\begin{tabular}[c]{@{}c@{}}LLM-based code modification,\\simulation, synthesis,\\verification-tool evasion\end{tabular} &
\begin{tabular}[c]{@{}c@{}}Security risks of AI-assisted\\hardware design\end{tabular} \\
\hline

HeLLO: CTF &
\begin{tabular}[c]{@{}c@{}}Logic-locked or obfuscated\\hardware IP provided for\\independent evaluation\end{tabular} &
\begin{tabular}[c]{@{}c@{}}Break the protection, recover\\functionality or keys, and\\report security weaknesses\end{tabular} &
\begin{tabular}[c]{@{}c@{}}Logic-locking attacks,\\structural analysis, functional\\recovery, key recovery, and\\obfuscation analysis\end{tabular} &
\begin{tabular}[c]{@{}c@{}}Independent evaluation of \\logic locking and hardware \\obfuscation\end{tabular} \\
\hline

WhibOx Contest &
\begin{tabular}[c]{@{}c@{}}Downloadable white-box AES \\or ECDSA implementations\\ containing embedded secret \\keys\end{tabular} &
\begin{tabular}[c]{@{}c@{}}Develop implementations\\resistant to key extraction or\\attack submitted implementations\\to recover their keys\end{tabular} &
\begin{tabular}[c]{@{}c@{}}Program analysis, reverse\\engineering, dynamic \\analysis,  cryptanalysis, \\implementation-specific attacks\end{tabular} &
\begin{tabular}[c]{@{}c@{}}Public developer-vs-attacker\\competition organized as a \\challenge\end{tabular} \\
\hline

\begin{tabular}[c]{@{}c@{}}HackTheSilicon\\@(DAC, DATE, CHES)\end{tabular} &
\begin{tabular}[c]{@{}c@{}}Open-box RTL SoC with\\specifications, firmware, and\\ security objectives\end{tabular} &
\begin{tabular}[c]{@{}c@{}}Find vulnerabilities, construct\\exploits, assess impact, and\\propose mitigations\end{tabular} &
\begin{tabular}[c]{@{}c@{}}RTL review, simulation, formal\\verification, static analysis,\\fuzzing, custom automation\end{tabular} &
\begin{tabular}[c]{@{}c@{}}Pre-silicon SoC security \\and security-aware EDA\end{tabular} \\
\hline
\end{tabular}%
}
\vspace{-1em}
\end{table*}

\subsubsection{Closed-Box vs. Open-Box Hardware CTFs}

Many hardware security competitions follow a closed-box model. Participants receive a physical device or development board but have little or no access to its internal implementation. They interact with the target through exposed ports, firmware, communication interfaces, or physical measurements. Typical methods include protocol analysis, fault injection, side-channel analysis, signal probing, firmware extraction, and hardware reverse engineering.
Closed-box competitions are useful for studying attacks against deployed hardware. They represent the view of an external attacker who has physical or logical access to a device. However, they provide limited support for studying the original RTL root cause. Participants may demonstrate that a device is vulnerable without being able to identify the exact design error that caused the weakness.
Hack@DAC was introduced as an open-box model in 2018. Participants received a vulnerable SoC at the RTL level together with supporting specifications and design artifacts. The task was to inspect the implementation and use simulation, formal verification, static analysis, emulation, or manual review to identify potential bugs. The main objective was not only to observe a failure but also to locate the root cause, determine its security impact, and propose a design-level correction. The organizers described this as a designer-centered approach to hardware hacking.

These two models are complementary. Closed-box competitions examine the security of implemented systems under realistic physical constraints. Open-box competitions examine whether security weaknesses can be detected and corrected before fabrication. Therefore, the exploration of hardware security benefits from both pre-silicon and post-silicon analysis.
Table \ref{tab:closed_open_ctfs} summarizes the key differences between closed-box and open-box hardware security competitions in terms of target access, analysis techniques, attacker perspective, validation evidence, and other related features.

\noindent $\bullet~$\textbf{\textit{Hackthesilicon Initiative}}: Hack@DAC was introduced in 2018 as an open-box hardware-security competition.
Following the first Hack@DAC event, the competition continued through later Design Automation Conference (DAC) editions. 
The HackTheSilicon event also records expansions to HACK@SEC, HACK@CHES, and HACK@DATE. HACK@SEC events were organized from 2020 to 2022, HACK@CHES events were held in 2021 and 2025, and HACK@DATE was introduced in 2025 \cite{hackthesilicon}.

\vspace{-1em}
\subsection{Comparing popular Hardware Security Competitions}

Community-driven hardware-security competitions provide a common platform for evaluating attacks, defenses, and analysis tools under shared assumptions. As summarized in Table \ref{tab:comparison}, these competitions differ mainly in their target abstraction and security objectives. The CSAW Logic Locking Conquest \cite{tan2020benchmarkingfrontierhardwaresecurity,sst_competition} and HeLLO: CTF \cite{helloctf} focus on the independent evaluation of Logic Locking, obfuscation and redaction techniques through key recovery, functional reconstruction, and structural attacks. The WhibOx Contest \cite{whibox_contest} applies a similar open-box model to cryptographic implementations, where attackers receive full access to protected AES or ECDSA implementations and attempt to recover embedded keys. The CSAW AI Hardware Attack Challenge \cite{csaw_ai_hardware_challenge} studies the security risks introduced by AI-assisted hardware design by requiring participants to insert and exploit hardware Trojans or design weaknesses. In contrast, HackTheSilicon competitions \cite{hackthesilicon}, including Hack@DAC, Hack@DATE, and Hack@CHES, use complete RTL-based SoCs and cover a broader range of security flaws, such as access-control violations, privilege-checking errors, and firmware–hardware interaction bugs. Together, these efforts support independent validation, expose weaknesses in existing evaluation methods, and promote more reproducible hardware-security research.

\begin{table*}[!ht]
\centering
\caption{Overview of the security-analysis strategies}
\label{tab:strategies}
\resizebox{\textwidth}{!}{%
\begin{tabular}{|c|c|l|l|l|l|}
\hline
\textbf{Strategy} &
  \textbf{Tools} &
  \multicolumn{1}{c|}{\textbf{Main objective}} &
  \multicolumn{1}{c|}{\textbf{Primary input}} &
  \multicolumn{1}{c|}{\textbf{Analysis output}} &
  \multicolumn{1}{c|}{\textbf{Validation evidence}} \\ \hline
\begin{tabular}[c]{@{}c@{}}Simulation-based \\ verification\end{tabular} &
  \begin{tabular}[c]{@{}c@{}}Synopsys VCS \\ / DVSim\end{tabular} &
  \begin{tabular}[c]{@{}l@{}}Identify failures under \\ executable test scenarios\end{tabular} &
  \begin{tabular}[c]{@{}l@{}}RTL, UVM environment, \\ DVSim tests, input seeds\end{tabular} &
  \begin{tabular}[c]{@{}l@{}}Failed tests, unexpected outputs, \\ abnormal state transitions\end{tabular} &
  \begin{tabular}[c]{@{}l@{}}Simulation logs and \\ waveform traces\end{tabular} \\ \hline
\begin{tabular}[c]{@{}c@{}}Formal property \\ verification\end{tabular} &
  Synopsys VC Formal &
  \begin{tabular}[c]{@{}l@{}}Check functional and security \\ properties exhaustively within   \\ the formal model\end{tabular} &
  \begin{tabular}[c]{@{}l@{}}RTL, SystemVerilog \\ Assertions\end{tabular} &
  \begin{tabular}[c]{@{}l@{}}Proven properties or \\ counterexamples\end{tabular} &
  \begin{tabular}[c]{@{}l@{}}Formal trace and \\ failed assertion\end{tabular} \\ \hline
\begin{tabular}[c]{@{}c@{}}Formal security \\ verification\end{tabular} &
  Synopsys VC Formal &
  \begin{tabular}[c]{@{}l@{}}Detect unauthorized information \\ flow between security domains\end{tabular} &
  RTL, Security Primitives &
  \begin{tabular}[c]{@{}l@{}}Confidentiality or integrity \\ violations\end{tabular} &
  \begin{tabular}[c]{@{}l@{}}Source-to-destination \\ propagation trace\end{tabular} \\ \hline
\begin{tabular}[c]{@{}c@{}}RTL Lint \\ analysis\end{tabular} &
  \begin{tabular}[c]{@{}c@{}}Synopsys VC Spyglass \\ Lint\end{tabular} &
  \begin{tabular}[c]{@{}l@{}}Detect structural and coding \\ issues that may cause   \\ security-relevant behavior\end{tabular} &
  \begin{tabular}[c]{@{}l@{}}RTL source code and \\ lint rules\end{tabular} &
  \begin{tabular}[c]{@{}l@{}}Warnings related to assignments, \\ widths, resets, and control logic\end{tabular} &
  \begin{tabular}[c]{@{}l@{}}Manual RTL review and \\ error logs\end{tabular} \\ \hline
\begin{tabular}[c]{@{}c@{}}LLM-assisted \\ bug detection\end{tabular} &
  Commercial LLMs &
  \begin{tabular}[c]{@{}l@{}}Identify bugs, CWE mappings, \\ and generate security properties\end{tabular} &
  \begin{tabular}[c]{@{}l@{}}RTL fragments, design \\ specifications, context\end{tabular} &
  \begin{tabular}[c]{@{}l@{}}Candidate bugs, CWE categories, \\ SVAs, and test ideas\end{tabular} &
  \begin{tabular}[c]{@{}l@{}}Conventional simulation or \\ formal verification\end{tabular} \\ \hline
\begin{tabular}[c]{@{}c@{}}Hybrid Fuzzing\end{tabular} &
  Synopsys VCS &
  \begin{tabular}[c]{@{}l@{}}Explore untested design states \\ and input sequences\end{tabular} &
  \begin{tabular}[c]{@{}l@{}}Random seeds, coverage\\database\end{tabular} &
  \begin{tabular}[c]{@{}l@{}}New tests, assertion violations, \\ and coverage improvements\end{tabular} &
  \begin{tabular}[c]{@{}l@{}}Coverage reports, logs, \\ and waveforms\end{tabular} \\ \hline
\end{tabular}%
}
\vspace{-1em}
\end{table*}

\begin{figure*}[!ht]
    \centering
    \includegraphics[width=0.7\textwidth]{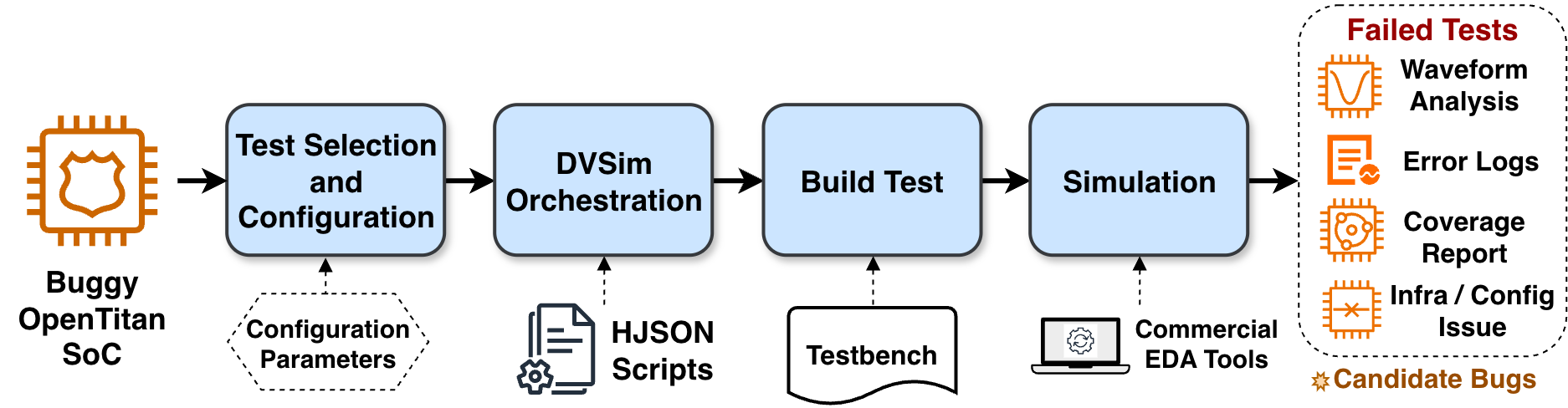}
    \caption{Illustration of the proposed DVSim-based simulation verification workflow.}
    \label{fig:dvsim_flow}
    \vspace{-1.5em}
\end{figure*}

\vspace{-1em}
\subsection{Role in Research, Training, and Tool Development}

Hardware-security competitions serve three main purposes. First, they provide practical training for participants to learn realistic SoC architectures, security requirements, RTL implementations, verification environments, and attack scenarios. This hands-on setting allows participants to develop practical skills in RTL analysis, simulation, formal verification, security-property development, exploit construction, and vulnerability reporting. 
Second, the competition artifacts provide evaluation benchmarks for research. 
Rather than evaluating a new technique only on small synthetic circuits or manually created bugs, researchers can use competition designs containing known security weaknesses and realistic SoC interactions. These benchmarks have supported the evaluation of diverse approaches, including information-flow tracking \cite{cellift,sec_ver_opentitan}, CWE-guided RTL analysis \cite{cweat_paper,spell}, assertion-based verification, LLM-assisted vulnerability detection \cite{nspg,lashed,lambda,marvel,lasso,atlas}, and symbolic fuzzing \cite{symbfuzz}. 
The availability of shared vulnerable designs, known bug sets, specifications, and verification infrastructure is particularly useful for security-aware EDA research. It enables researchers to measure whether a proposed method can identify known vulnerabilities, analyze different classes of weaknesses, study false positives, and compare complementary verification strategies on a common target. Third, the competitions support interaction between academia and industry. Academic teams can test new analysis methods on realistic designs. Industry participants can study weakness classes that may not be covered by traditional functional verification. Tool vendors can evaluate whether existing design automation tools support security-specific use cases.

\section{Security Analysis Methodologies}
\label{sec:methodology}

To address the diverse categories of vulnerabilities, we adopted a multi-strategy methodology for security analysis that combines simulation, formal verification, structural analysis, LLM-assisted reasoning, fuzzing, and exploit development. Each strategy generates candidate vulnerabilities from a different view of the design. The candidates are then validated using independent evidence before they are reported.
The methodologies were applied during Hack@DAC (2024,2025) and Hack@CHES (2025) competitions to the buggy OpenTitan SoC design. The provided infrastructure included RTL source code, design verification environments, security properties, simulation tests, and commercial EDA tools. The security verification analysis strategies are summarized in Table \ref{tab:strategies}.

\vspace{-1em}
\subsection{Analysis Workflow}

The methodology consists of four main phases. First, the design environment is built and the available test, formal, and lint flows are executed to establish a baseline. Second, each analysis strategy is used to generate vulnerability candidates. Third, the candidates are reviewed to remove tool errors, invalid assumptions, and duplicate findings. Finally, confirmed vulnerabilities are reproduced using a custom test or formal trace, mapped to a relevant hardware CWE, and documented with their security impact and possible mitigation.
The strategies are not treated as independent experiments. Results from one method guide the use of another method. For example, a failing regression test may identify a suspicious state transition that is later expressed as an assertion. A lint warning may identify a width or assignment error that is validated through simulation. An LLM-generated security property may be checked using formal verification. This cross-validation reduces false positives and provides stronger evidence for each reported vulnerability.

\vspace{-1em}
\begin{table*}[!hb]
\centering
\caption{Representative vulnerabilities identified using simulation-based security verification.}
\label{tab:simulation_bugs}
\resizebox{\textwidth}{!}{%
\begin{tabular}{p{0.10\textwidth} p{0.30\textwidth} p{0.20\textwidth} p{0.31\textwidth}}
\hline
\textbf{Target IP} & \textbf{Security Weakness} & \textbf{RTL Location} & \textbf{Observed Impact} \\
\hline
AES &
Sensitive state and key registers are not properly wiped during the security-clear operation. &
\texttt{aes\_cipher\_core.sv} &
Sensitive AES state or key information may remain exposed instead of being cleared. \\

CSRNG &
The default state handling the CSRNG read-data return is missing. &
\texttt{csrng\_reg\_top.sv} &
Correct read data, error codes, or alert signals may not be generated as required. \\

LC\_CTRL &
DFT and debug functionality are enabled in the \texttt{LcStProd} state. &
\texttt{lc\_ctrl\_signal\_decode.sv} &
Debug and DFT functionality can remain accessible during the production lifecycle state. \\

OTBN &
Secure-wipe control permits \texttt{secure\_wipe\_req\_o} to be asserted after a wipe operation has already started. &
\texttt{otbn\_controller.sv} &
The secure-wipe control flow may enter an inconsistent state and result in incomplete or incorrect wiping. \\

KMAC &
The sparse-FSM error indication is delayed until an inserted counter exceeds a threshold. &
\texttt{kmac\_core.sv} &
Error detection is delayed, allowing incorrect operation before the sparse-FSM error is reported. \\
\hline
\end{tabular}%
}
\end{table*}

\subsection{Simulation-Based Security Verification}

Dynamic simulation-based verification aims to identify functional deviations and security-relevant behaviors by utilizing regression test suites to systematically evaluate target IPs. 
The strategy utilizes Synopsys VCS combined with the DVSim framework available with the existing OpenTitan \cite{opentitan} design verification environment and custom automation scripts. Smoke tests and regression tests were run across the selected IP blocks. The goal was to identify failing tests, unexpected output values, assertion failures, timeouts, and abnormal state transitions.
Fig.~\ref{fig:dvsim_flow} illustrates the automated DVSim-based verification workflow, including test configuration, regression execution, failure detection, and analysis.

We developed an HJSON-based orchestration layer to systematically execute the existing DVSim tests across selected SoC IPs. For each IP, the configuration specifies the native DVSim configuration file, individual tests or regressions to execute (set \texttt{true} or \texttt{false}), and fixed simulation seeds for reproducibility. The wrapper invokes DVSim for each enabled test and preserves the corresponding simulation results for subsequent failure analysis. DVSim internally performs the required build and simulation stages and supports independently seeded test instances and regression-level reporting. Listing~\ref{lst:dvsim_hjson} presents the generic HJSON-based test orchestration configuration used to specify the target IP, selected tests or regressions, and simulation seeds for automated execution.

\begin{lstlisting}[
    style=hjsonstyle,
    caption={Generic HJSON configuration for automated simulation-based
    security verification.},
    label={lst:dvsim_hjson}
]
{
    repo_root: "<PATH_TO_OPENTITAN_DIRECTORY>"
    // Design under evaluation
    design_variant: "buggy"
    // Simulator used by the DV environment
    simulator: "vcs"
    // IPs selected for security analysis
    test_ip_list: [
        {
            name: "<IP_NAME>" // e.g. hmac
            sim_cfg: "hw/ip/<IP_NAME>/dv/\
                        <IP_NAME>_sim_cfg.hjson"
            has_mask: false  // Optional
            test_list: [
                {
                    type: "regression"
                    name: "smoke"
                    skip: true/false
                }
                {
                    type: "run_test"
                    name: "<IP_NAME>_smoke"
                    skip: true/false
                }
                {
                    type: "run_test"
                    name: "<SECURITY_TEST_1>"  
                    //name (e.g. hmac_wipe_secret)
                    skip: false
                    seeds: [101, 102, 103]
                }
            ]
        }
    ]
}
\end{lstlisting}

The execution sequence consists of the following phases:

\begin{enumerate}
    \item Exhaustive Smoke Regression: Automated scripts run standard regression smoke tests across all SoC IP blocks to isolate failing test instances.  
    \item Configuration Automation: An automation wrapper file (.hjson) dynamically configures the simulation parameters, adjusting tests across specific target IPs.  
    \item Log File Diagnostics: Analysis of runtime logs determines the precise cycles where the design under test (DUT) deviates from expected architectural behaviors. 
\end{enumerate}

The failed cases that were examined using simulation logs and traces may not be directly classified as vulnerabilities unless they violate defined security requirements. The associated logs, assertions, scoreboard results, and waveform traces are analyzed to determine the RTL root cause and whether the observed behavior violates a defined security requirement. For example, a test failure related to a protected register was considered security relevant only when an unauthorized component could update or observe protected data.
This distinction was important because regression failures can also result from incorrect constraints, unsupported configurations, or testbench errors. As summarized in Table~\ref{tab:simulation_bugs}, simulation exposed vulnerabilities involving incomplete secret wiping, incorrect FSM behavior, improperly enabled debug functionality, and delayed error handling. These bugs were identified through failed DVSim smoke, stress, and security-countermeasure tests, followed by RTL-level analysis of the corresponding failure conditions.

\begin{table*}[t]
\centering
\caption{Representative vulnerabilities identified using formal verification techniques.}
\label{tab:formal_bugs}
\resizebox{\textwidth}{!}{%
\begin{tabular}{p{0.09\textwidth} p{0.10\textwidth} p{0.29\textwidth} p{0.19\textwidth} p{0.27\textwidth}}
\hline
\textbf{Target IP} & \textbf{Method} & \textbf{Security Weakness} & \textbf{RTL Location} & \textbf{Observed Impact} \\
\hline

AES &
FSV &
Key-share and input-data values are propagated to the software-visible register read path. &
\texttt{aes\_reg\_top.sv} &
Secret key and plaintext-related data can be exposed through unauthorized register reads. \\

HMAC &
FPV &
\texttt{wipe\_secret\_we} is asserted under an incorrect dependency on \texttt{reg\_error}. &
\texttt{hmac\_reg\_top.sv} &
Internal secret registers can be wiped under an unintended register-error condition. \\

HMAC &
FPV &
HMAC key registers \texttt{reg2hw.key[0:2]} are assigned to \texttt{reg\_rdata\_next}. &
\texttt{hmac\_reg\_top.sv} &
The HMAC hashing key can be exposed through the register read-data path. \\

UART &
FPV &
The receiver parity-error signal does not depend on \texttt{parity\_enable}. &
\texttt{uart\_rx.sv} &
Incorrect parity-error reporting affect FIFO sync and generate false interrupts. \\
\hline
\end{tabular}%
}
\vspace{-1.5em}
\end{table*}

\begin{table*}[!hb]
\centering
\caption{Representative vulnerabilities identified using RTL lint and structural analysis.}
\label{tab:lint_bugs}
\resizebox{\textwidth}{!}{%
\begin{tabular}{p{0.12\textwidth} p{0.29\textwidth} p{0.19\textwidth} p{0.31\textwidth}}
\hline
\textbf{Target IP} & \textbf{Security Weakness} & \textbf{RTL Location} & \textbf{Observed Impact} \\
\hline
HMAC &
The secret-wipe control depends incorrectly on \texttt{reg\_error}. &
\texttt{hmac\_reg\_top.sv} &
Internal secret registers can be wiped under an unintended register-error condition. \\

Primitive Shadow Register &
\texttt{err\_update} and \texttt{err\_storage} depend on \texttt{error\_s}, although \texttt{error\_s} is not initialized or updated. &
\texttt{prim\_subreg\_shadow.sv} &
The intended error-detection logic can operate incorrectly. \\

Power Manager &
The \texttt{rom\_intg\_chk\_good} check is bypassed in \texttt{FastPwrStateRomCheckGood}, and the default terminal-state handling is removed. &
\texttt{pwrmgr\_fsm.sv} &
ROM integrity checking is disabled under the affected lifecycle test condition. \\
\hline
\end{tabular}%
}
\end{table*}

\vspace{-1em}
\subsection{Formal Property and Security Verification}

Formal Property Verification (FPV) is used to check whether the selected functional and security requirements hold for all behaviors represented by the formal model, whereas Formal security verification (FSV) is used to analyze unintended information flow from security-sensitive sources to unauthorized destinations.
For FPV, security requirements were translated into assertions describing the expected behavior of the target RTL. The properties were formally checked against the design, and failed properties were analyzed using the generated counterexamples to identify the corresponding RTL conditions. This analysis exposed weaknesses involving secret-wipe control, unauthorized register access, and incorrect parity-error handling.
FSV was applied to security-sensitive data paths by defining protected sources and unauthorized destinations and checking whether information could propagate between them. This analysis identified an unintended path exposing AES key shares and input-data values through the software-visible register read interface.

Table~\ref{tab:formal_bugs} summarizes representative vulnerabilities identified using FPV and FSV. FPV exposed incorrect security-control dependencies and register-access behavior through assertion checking, while FSV identified an unauthorized information-flow path from security-sensitive AES registers to a software-visible destination. These findings demonstrate the complementary use of property-based and information-flow-oriented formal analysis for detecting control and confidentiality-related weaknesses.

\subsection{RTL Lint and Syntax Checking}

Structural RTL analysis was performed using Synopsys VC SpyGlass Lint. The design IPs were built using the available lint flow, and the resulting warnings and errors were reviewed for possible security relevance.
Lint analysis can identify several implementation patterns that may contribute to vulnerabilities, including incomplete assignments and unintended latch inference, width mismatches, constant or unreachable conditions, uninitialized registers, multiple signal drivers, incorrect reset values, truncated arithmetic operations, unused control signals, inconsistent case statements, and combinational loops.

Most lint warnings are not security vulnerabilities by themselves. Each warning was therefore analyzed in the context of the affected module and its security function. Higher priority was given to warnings involving key registers, privilege checks, error logic, access-control signals, state machines, and externally visible outputs.
A security-relevant lint finding was confirmed through RTL inspection and, where possible, simulation or formal verification. This step prevented general coding-style warnings from being reported as vulnerabilities.

Lint analysis provides fast coverage across many modules and is useful for identifying simple but high-impact design errors. However, lint analysis lacks system-level security context, as structural analysis alone cannot determine whether a suspicious signal relationship is exploitable. Table~\ref{tab:lint_bugs} includes examples involving incorrect error-signal usage, secret-wipe control, and missing integrity-related FSM logic.

\vspace{-1em}
\subsection{LLM-Assisted Vulnerability Analysis}

\begin{figure*}[!ht]
    \centering
    \includegraphics[width=0.9\textwidth]{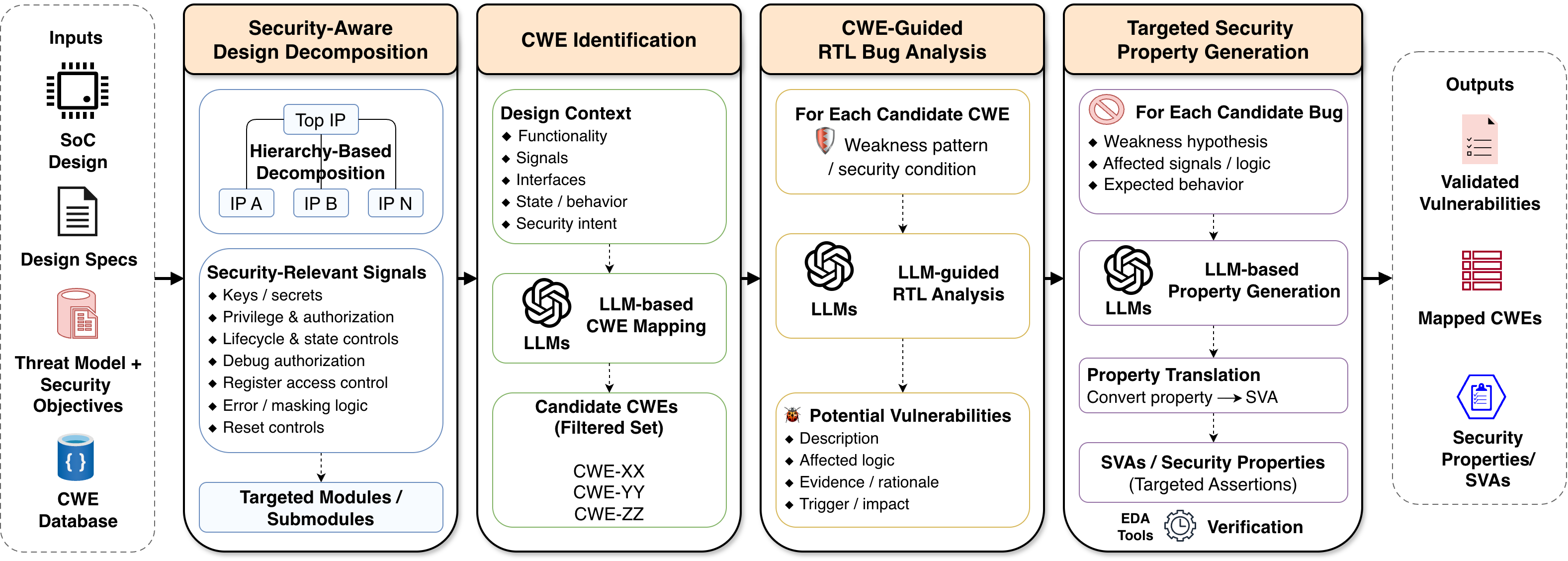}
    \caption{Flow diagram of the proposed LLM-assisted vulnerability-analysis workflow.}
    \label{fig:llm_flow}
\end{figure*}

LLMs have been increasingly applied to hardware design and verification tasks, including natural-language specification interpretation, HDL generation, RTL analysis, and verification automation \cite{llm_survey}. Their adoption has enabled applications such as vulnerability analysis, CWE mapping, security property generation, and bug localization \cite{Paria2026}. However, directly applying LLMs to large SoCs remains challenging due to design complexity, context limitations, and potentially incorrect or unsupported outputs. The proposed LLM-assisted analysis focuses on identifying candidate bugs and related CWEs and validating security properties using conventional EDA tools, as depicted in Fig. \ref{fig:llm_flow}.

\begin{table*}[t]
\centering
\caption{Representative vulnerabilities analyzed using the LLM-assisted approach.}
\label{tab:llm_bugs}
\resizebox{\textwidth}{!}{%
\begin{tabular}{p{0.10\textwidth} p{0.30\textwidth} p{0.20\textwidth} p{0.31\textwidth}}
\hline
\textbf{Target IP} & \textbf{Security Weakness} & \textbf{RTL Location} & \textbf{Observed Impact} \\
\hline

AES &
The \texttt{SecAllowForcingMasks} parameter is hardcoded to 1, disabling the intended masking protection. &
\texttt{aes.sv:180} &
The masking countermeasure can be disabled, resulting in the reported side-channel vulnerability. \\

HMAC &
\texttt{wipe\_secret\_we} is asserted when \texttt{reg\_error} is active. &
\texttt{hmac\_reg\_top.sv:2121} &
Internal secret registers can be wiped under an unintended register-error condition. \\

HMAC &
The \texttt{reg2hw.key\_*} values are propagated to \texttt{reg\_rdata\_next}. &
\texttt{hmac\_reg\_top.sv:2419, 2423, 2623} &
A malicious register read can expose the HMAC secret key. \\

\hline
\end{tabular}%
}
\vspace{-1em}
\end{table*}

\subsubsection{Security-Aware Design Decomposition}
The first step limits the analysis to security-relevant regions of the SoC. Each target IP is decomposed into modules or submodules based on the design hierarchy. The analysis prioritizes modules containing security-sensitive assets or controls, such as cryptographic keys, privilege signals, lifecycle states, debug authorization, register access controls, integrity checks, masking logic, reset controls, and error-handling signals.
Table \ref{tab:llm_context} summarizes the context provided during LLM-guided analysis.

\subsubsection{CWE Identification}
The next step identifies hardware CWEs that are relevant to the functionality of the selected module. Instead of comparing every RTL module against the complete CWE database, the design context is used to narrow the candidate weakness classes. For example, a cryptographic block may be analyzed for weaknesses related to the protection of secret data, masking, key handling, or unauthorized access, while a lifecycle controller may require greater emphasis on state-transition and privilege-control weaknesses.

\subsubsection{CWE-Guided Bug Analysis}
The identified CWEs are used to guide targeted RTL analysis. For each candidate CWE, the LLM is prompted to determine whether the corresponding weakness pattern appears in the selected RTL region. Typical checks include missing or incorrect access-control checks, improper dependencies on authorization or error signals, unintended propagation of protected data, disabled cryptographic countermeasures, insecure hardcoded parameters, invalid FSM transitions, unauthorized modification of secret registers, and security violations triggered by reset, error, or exceptional conditions.

\subsubsection{Targeted Security-Property Generation}
For each candidate bug, the LLM generates a security property that expresses the expected behavior of the affected logic. The property is generated after a candidate weakness has been identified. Therefore, the assertion is targeted toward confirming a specific hypothesis instead of broadly generating properties for the entire module. The generated properties are translated into equivalent SVAs. Existing studies \cite{lasa,divas} show that LLM-generated assertions may contain incorrect clocking, reset semantics, signal references, or temporal operators ; therefore, every generated SVA has been validated syntactically and semantically.

Table~\ref{tab:llm_bugs} presents representative vulnerabilities analyzed using the LLM-assisted workflow with pre-trained GPT-4o as the baseline LLM. The analysis identified security-relevant RTL patterns involving disabled cryptographic masking, incorrect secret-wipe control, and the propagation of secret key values to a software-visible register path. These findings were used to generate targeted security properties that could subsequently be validated using conventional simulation or formal verification tools.

\begin{table}[!htbp]
\centering
\caption{Context Provided for LLM-Guided Bug Analysis}
\label{tab:llm_context}
\renewcommand{\arraystretch}{1.25}
\resizebox{\columnwidth}{!}{
\begin{tabular}{p{2.5cm}p{5.8cm}}
\toprule
\textbf{Context} & \textbf{Information Provided} \\
\midrule
Design hierarchy &
Target IP, module, submodule, and interfaces \\ 

\multirow{2}{*}{Security assets} &
Keys, protected registers, privilege states, configuration data,
and security-critical control signals \\ 

\multirow{2}{*}{Security intent} &
Expected access rules, legal state transitions, reset behavior,
and confidentiality/integrity requirements \\ 

RTL slice &
Relevant control/data logic and dependent signals \\ 

\multirow{2}{*}{Interface context} &
Register map, bus signals, privilege information, interrupts,
alerts, and debug interfaces \\ 

\multirow{2}{*}{Verification evidence} &
Failed test logs, assertion failures, lint warnings,
counterexamples, or waveform observations \\ 

Weakness knowledge &
Relevant CWE descriptions and common patterns \\ 
\bottomrule
\end{tabular}}
\vspace{-1em}
\end{table}

\begin{table*}[!ht]
\centering
\caption{Representative vulnerabilities identified using fuzzing-based analysis.}
\label{tab:fuzz_bugs}
\resizebox{\textwidth}{!}{%
\begin{tabular}{p{0.13\textwidth} p{0.35\textwidth} p{0.12\textwidth} p{0.35\textwidth}}
\hline
\textbf{Target IP} & \textbf{Security Weakness} & \textbf{RTL Location} & \textbf{Observed Impact} \\
\hline
Entropy Source &
Both \texttt{test\_fail\_hi\_pulse\_o} and \texttt{test\_fail\_lo\_pulse\_o} are forced to constant values, bypassing the intended failure signaling. &
\begin{tabular}[c]{@{}c@{}}\texttt{entropy\_src\_}\\\texttt{markov\_ht.sv}\end{tabular} &
Health-test failure signaling is bypassed, affecting detection of randomized-threshold violations. \\

UART &
The trigger logic is modified such that triggering occurs unconditionally. &
\texttt{uart\_core.sv} &
Unintended execution, glitches, or incorrect interrupt generation can occur. \\

SPI TPM &
The locality check uses an incorrect comparison with \texttt{NumLocality}. &
\texttt{spi\_tpm.sv} &
The \texttt{invalid\_locality} condition can be set incorrectly. \\
\hline
\end{tabular}%
}
\vspace{-1.5em}
\end{table*}

\subsection{Hybrid Fuzzing}

\begin{figure}[!htbp]
    \centering
    \includegraphics[width=0.725\columnwidth]{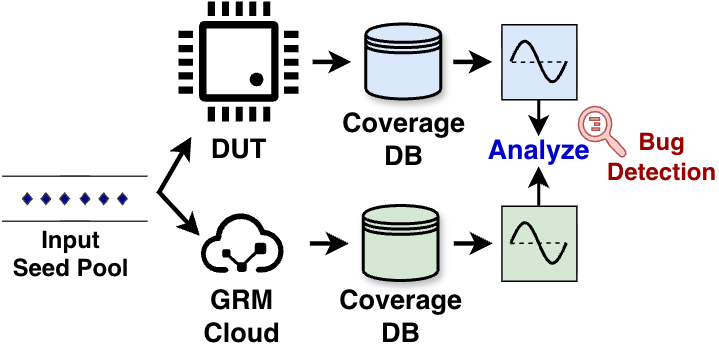}
    \caption{Coverage-guided Fuzzing approach.}
    \label{fig:cgf}
    \vspace{-1.5em}
\end{figure}

Coverage-guided fuzzing (CGF) is used to systematically verify hardware designs and quantify verification completeness \cite{fuzzing_survey,fuzzody}. In this methodology, a hardware fuzzer automatically generates and mutates input sequences (e.g., instruction streams, transaction sequences, or test vectors) and executes them on the design-under-test (DUT). During execution, coverage metrics information \cite{rscovglsvlsi} is collected and stored in a coverage database, as shown in Fig.~\ref{fig:cgf}. This feedback guides the fuzzer toward unexplored design behaviors by prioritizing inputs that exercise new states, control paths, or functional regions. The primary objective is to maximize exploration of
the code coverage of the hardware and expose functional errors or security-critical vulnerabilities that may remain undetected through conventional verification approaches.

\begin{figure}[!htbp]
    \centering
    \includegraphics[width=0.675\columnwidth]{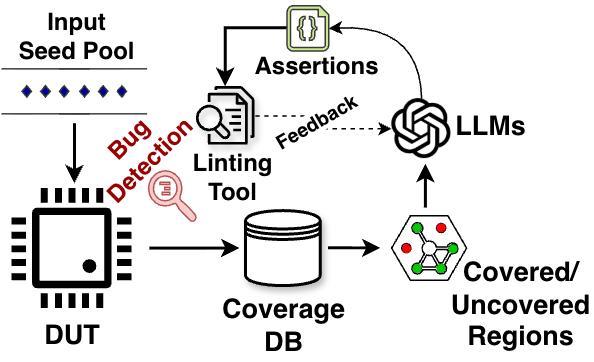}
    \caption{Hybrid fuzzing approach with feedback-driven exploration of uncovered hardware regions.}
    \label{fig:hybrid}
    \vspace{-1em}
\end{figure}

Unlike conventional coverage-guided fuzzers that rely primarily on mutation strategies to explore uncovered areas, the proposed hybrid fuzzing first analyzes the coverage database to identify regions that remain unexplored or are difficult to activate. These uncovered coverage points are then provided as context to the LLM, which generates targeted verification artifacts, such as SVAs, directed test scenarios, or input constraints designed to exercise the corresponding design behaviors. The generated assertions or constraints are subsequently analyzed by verification tools to ensure syntactic correctness and eliminate invalid specifications. The validated artifacts are then incorporated into the fuzzing framework to guide input generation toward the previously uncovered regions. Fig.~\ref{fig:hybrid} illustrates the integration of LLMs into the hardware fuzzing loop to improve the exploration of hard-to-reach design regions. The representative findings in Table~\ref{tab:fuzz_bugs} include incorrect UART triggering and faulty TPM locality checking in SPI, illustrating the usefulness of input-driven exploration for exposing abnormal behavior.

\vspace{-1em}
\subsection{Custom Exploit Development}

The exploit development process began with the reported bug description, which provided information including the affected module, vulnerable code location, triggering conditions, expected behavior, security impact, and proposed mitigation. This information, together with the OpenTitan \cite{opentitan} project directory hierarchy, was provided to the LLM to identify the RTL and verification components associated with the reported vulnerability. After identifying the relevant components, the corresponding RTL and verification source files were retrieved and supplied to the LLM for further analysis. Based on the bug description and the implementation of the identified components, the LLM generated the modifications required to reproduce the vulnerability while preserving the original design behavior. Depending on the reported vulnerability, the generated changes included extending an existing UVM verification sequence, introducing a lightweight custom UVM sequence, adding verification-side monitoring mechanisms such as SystemVerilog Assertions (SVAs), or incorporating functional checks to observe the exploit outcome.

The generated modifications were integrated into the OpenTitan \cite{opentitan} project and executed using the existing UVM-based verification environment. The exploit execution was validated by analyzing simulation outputs, including assertion failures, register state changes, interrupt status, protocol responses, and simulation logs, to confirm that the reported vulnerability was successfully reproduced.

\subsubsection{Application to the UART \texttt{lsio\_trigger\_o} Exploit}

For the UART vulnerability, the bug description and project hierarchy were provided to the LLM to identify the components involved in the DMA trigger generation logic. The corresponding RTL and verification files were then supplied to the LLM, which modified the existing \texttt{uart_fifo_reset_vseq.sv} sequence to generate the FIFO conditions required to trigger the reported vulnerability and generated an SVA to automatically detect incorrect assertions of \texttt{lsio_trigger_o}. The generated sequence also included \texttt{UVM_INFO} messages that report the executed test case and FIFO levels, providing the operating conditions under which each assertion failure occurred. The resulting simulation output is shown in Listing~\ref{lst:uart-output}.

\begin{lstlisting}[
style=SimOutput,
caption={UART \texttt{lsio_trigger_o} exploit simulation output},
label={lst:uart-output}
]
UVM_INFO uart_fifo_reset_vseq.sv: Test 1: Sending 5 TX bytes below watermark
Error: Assertion LsioTriggerInvalid_A failed: lsio_trigger_o is 1 when TX FIFO (5) >= 8 or RX FIFO (0) < 16
UVM_INFO uart_fifo_reset_vseq.sv: TX FIFO level: 5, RX FIFO level: 0

UVM_INFO uart_fifo_reset_vseq.sv: Test 2: Sending 10 RX bytes below watermark
Error: Assertion LsioTriggerInvalid_A failed: lsio_trigger_o is 1 when TX FIFO (0) >= 8 or RX FIFO (10) < 16
UVM_INFO uart_fifo_reset_vseq.sv: TX FIFO level: 0, RX FIFO level: 10

UVM_INFO uart_fifo_reset_vseq.sv: Test 3: Checking lsio_trigger_o after FIFO operations
Error: Assertion LsioTriggerInvalid_A failed: lsio_trigger_o is 1 when TX FIFO (0) >= 8 or RX FIFO (0) < 16
UVM_INFO uart_fifo_reset_vseq.sv: TX FIFO level: 0, RX FIFO level: 0
\end{lstlisting}

As shown in Listing~\ref{lst:uart-output}, the assertion failures indicate that \texttt{lsio_trigger_o} remained asserted under invalid FIFO conditions, while the accompanying \texttt{UVM_INFO} messages identify the FIFO occupancy that produced each failure.

\subsubsection{Application to the HMAC \texttt{WIPE\_SECRET} Exploit}

For the HMAC vulnerability, the bug description and project hierarchy were provided to the LLM to identify the RTL and verification components associated with the \texttt{WIPE_SECRET} register write logic. After analyzing the relevant source files, the LLM generated a lightweight custom UVM sequence to configure the HMAC module, issue the malformed TL-UL write transaction, and perform functional checks on the resulting digest registers and interrupt status. The generated sequence also included \texttt{UVM_INFO} messages that report the execution stages together with the observed digest values and interrupt status used by the functional checks. The resulting simulation output is shown in Listing~\ref{lst:hmac-output}.

\begin{lstlisting}[
style=SimOutput,
caption={HMAC \texttt{WIPE_SECRET} exploit simulation output},
label={lst:hmac-output}
]
UVM_INFO @ <time>: SEQ [SEQ] DIGEST_7: 0x<valid_digest_7>
UVM_INFO @ <time>: SEQ [SEQ] Triggering buggy WIPE_SECRET write...
UVM_INFO @ <time>: SEQ [SEQ] SUCCESS: HmacErr interrupt triggered.
UVM_INFO @ <time>: SEQ [SEQ] DIGEST after buggy write:
UVM_INFO @ <time>: SEQ [SEQ] DIGEST_0: 0xffffffff
UVM_INFO @ <time>: SEQ [SEQ] DIGEST_1: 0xffffffff
\end{lstlisting}

As shown in Listing~\ref{lst:hmac-output}, the reported digest values and \texttt{HmacErr} interrupt confirm that the malformed transaction successfully reproduced the reported vulnerability.

\begin{figure}[!htbp]
    \centering
    \subfloat[]{\label{fig:bug_classes}
    \includegraphics[width=\columnwidth]{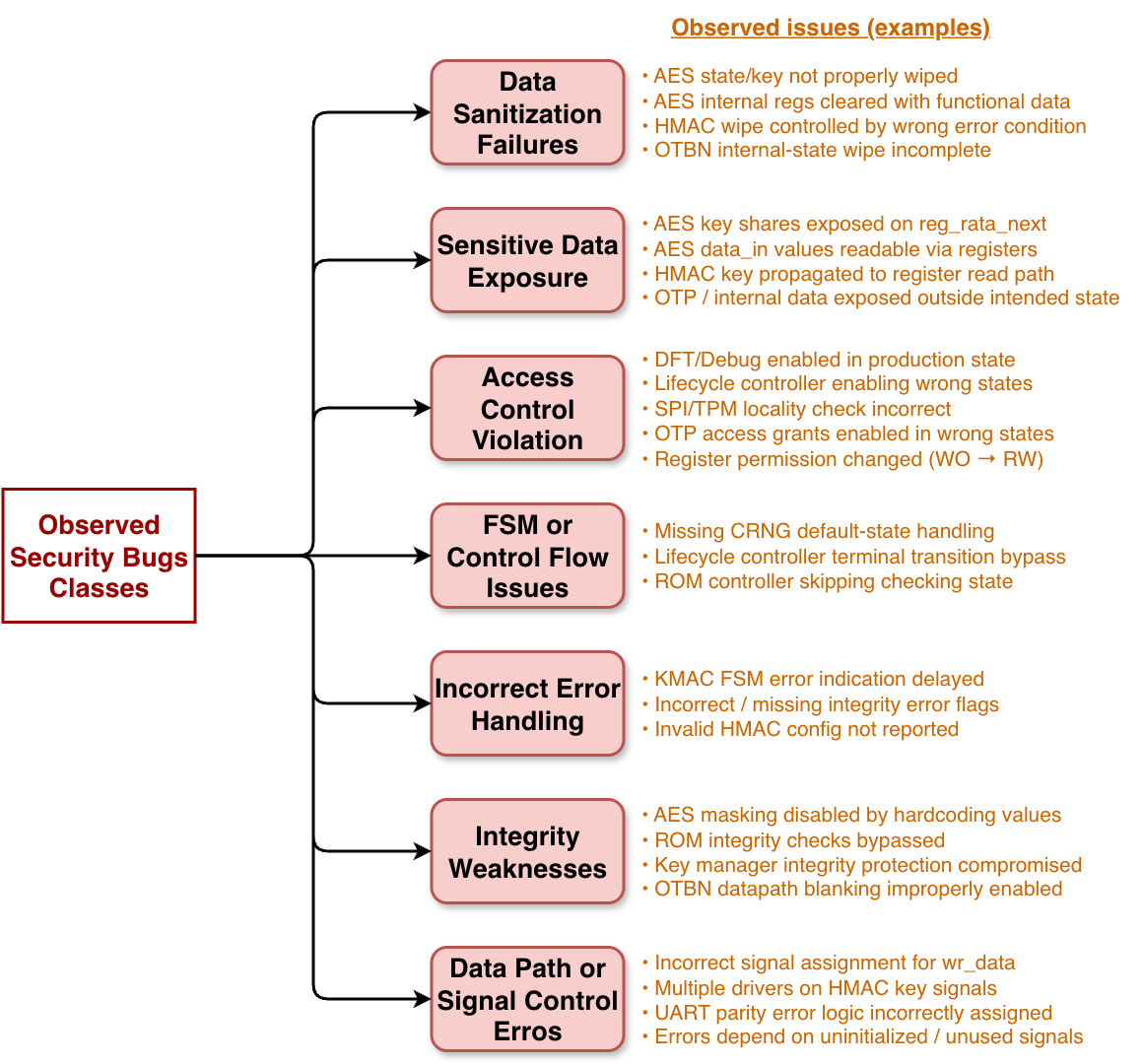}}
    \hfill
    \subfloat[]{\label{fig:bug_dist}
    \includegraphics[width=0.75\columnwidth]{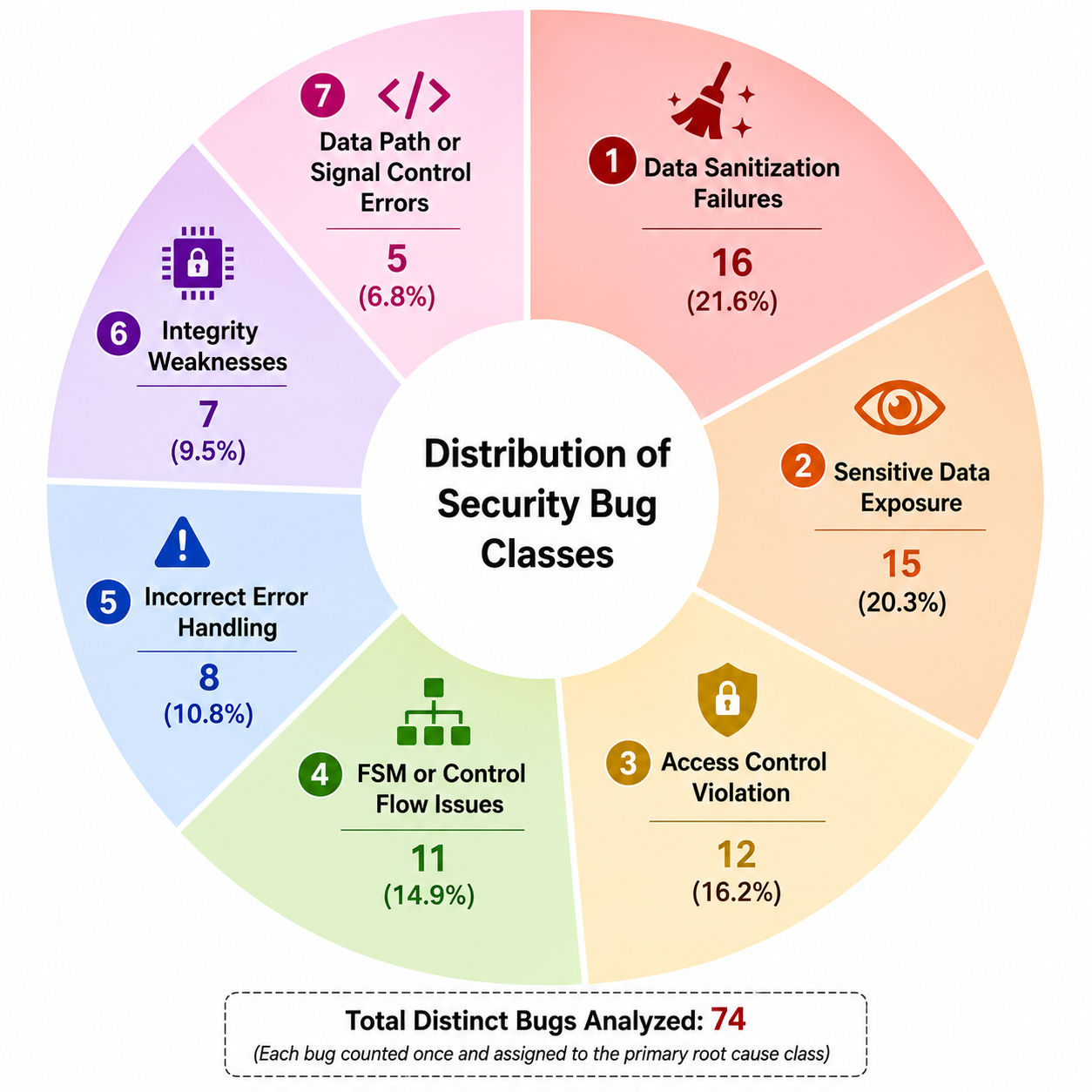}}
    \caption{(a) Taxonomy of observed security bugs; (b) distribution of the identified bug classes.}%
    \label{fig:bug_class_dist}
    \vspace{-1em}
\end{figure}

\vspace{-1em}
\subsection{Bug Classes and Distribution}

The reported vulnerabilities were grouped into seven security classes based on their primary root cause and security impact: data sanitization failures, sensitive data exposure, access-control violations, FSM/control-flow issues, incorrect error handling, integrity weaknesses, and data-path/signal-control errors.
Fig.~\ref{fig:bug_classes} illustrates the resulting taxonomy, while Fig.~\ref{fig:bug_dist} summarizes the distribution. The classes provide an analysis-oriented grouping of the observed bugs and do not replace their individual CWE mappings. Each distinct bug was assigned once to the class that best represented its primary RTL root cause to avoid double counting across overlapping security effects. Data sanitization failures and sensitive data exposure represent the largest fractions, followed by FSM/control-flow issues and incorrect error handling. Access-control violations, integrity weaknesses, and data-path/signal-control errors account for the remaining findings. The distribution indicates that a substantial portion of the observed bugs originated from the incorrect handling of sensitive state, security-critical control logic, and error conditions rather than from isolated syntax-level defects.

The taxonomy also highlights the diversity of security failures encountered across the competitions. The findings span confidentiality-related leakage, improper secret clearing, lifecycle and privilege violations, faulty state transitions, weakened countermeasures, and incorrect RTL signal dependencies. This diversity further supports the need for complementary verification techniques, as different bug classes are exposed more effectively by different forms of simulation, formal analysis, linting, fuzzing, and LLM-assisted inspection.

\section{Lessons Learned}
\label{sec:lessons}

The following subsections summarize the key lessons related to security intent, complementary verification, cross-layer analysis, and reproducible vulnerability validation.

\subsubsection{Security Intent Is Critical for Bug Identification}

A functional failure does not necessarily indicate a security vulnerability. The analysis must include the protected assets, trust boundaries, privilege rules, and expected security behavior before interpreting a failed test or RTL code as a security vulnerability. Several important bugs arise from logic that is functionally valid under normal operation but violates a security requirement under unauthorized, error, or exceptional conditions. Therefore, security analysis should begin with identifying the intended protection mechanisms and then checking whether the RTL correctly enforces them.

\subsubsection{Combine Complementary Verification Methods}

No single verification method provides sufficient coverage for all SoC security weaknesses. Simulation is effective for reproducing executable attack sequences, while formal verification is better suited for temporal properties and hard-to-reach states. Static analysis provides scalable structural screening, and information-flow analysis can directly examine confidentiality and integrity violations \cite{hardfails,hackdac_2021,cellift,sec_ver_opentitan}. A practical direction is therefore to develop hybrid verification flows in which findings from one technique automatically guide another, rather than applying each method independently.

\subsubsection{Verification Infrastructure Requires Security-Aware Guidance}

Existing regression environments, assertions, testbenches, and lint flows provide a useful starting point for vulnerability discovery. However, functional regressions mainly exercise expected design behavior and may not target unauthorized access, malicious transitions, or abnormal error conditions. Lint-based analysis can be combined with manual inspection and targeted tests when analyzing SoC security bugs \cite{bug_hunting_1,bug_hunting_2}. An important direction is therefore to augment existing verification infrastructure with security-directed stimulus generation and coverage metrics that measure security-relevant states and behaviors rather than functional coverage alone.

\subsubsection{Automated Findings Require Independent Validation}

Lint warnings, LLM-generated bug hypotheses, failed assertions, and fuzzing anomalies should be treated as vulnerability candidates rather than confirmed bugs. Each candidate should be traced to an RTL root cause and reproduced through a simulation trace, formal counterexample, information-flow path, or custom exploit. This is particularly important for LLM-assisted analysis, where generated properties can be syntactically valid but semantically incorrect \cite{lasa,Paria2026}. A useful direction is, therefore, closed-loop AI-assisted verification, where LLM reasoning generates candidates while simulation or formal tools provide the acceptance criterion.

\subsubsection{Cross-Layer Analysis Is Necessary for SoC-Level Bugs}

Many security vulnerabilities cannot be detected by analyzing isolated RTL modules. Their activation may depend on register transactions, firmware operations, privilege states, or translation to gate-level netlists \cite{hardfails,synfuzz}. Future verification techniques should therefore support cross-module verification across different abstraction levels.

\subsubsection{Tool Automation and Parallelization Improve Coverage}

Large SoCs contain many IPs and security-critical interfaces, making exhaustive manual analysis impractical under the time constraints during competitions. Automation of test execution, log collection, property checking, and result triage reduces repetitive effort. A coordinated multi-method agentic LLM-driven workflow \cite{marvel,agentic_vts,ver_agentic_1} provides better coverage than sequential analysis by a single technique.

\section{Broader Impact and Research Directions}
\label{sec:impact}

Hardware hacking competitions provide benefits beyond vulnerability discovery by creating shared platforms for security training, tool evaluation, and benchmark development. Their open-box structure allows participants from hardware design, verification, firmware, and security backgrounds to analyze the target SoC under common threat models and evaluation criteria. This promotes cross-domain collaboration and exposes participants to practical security-analysis workflows that are difficult to reproduce through isolated classroom or laboratory exercises.

\subsubsection{Security Training and Community Development}
These competitions provide hands-on experience in threat modeling, RTL analysis, simulation, formal verification, exploit development, and vulnerability reporting. The interaction between participants with different technical backgrounds also supports knowledge transfer across hardware design, verification, and security domains. Such exposure is particularly useful for developing security-aware design and verification skills and for training researchers entering the hardware-security field.

\subsubsection{Reusable Benchmarks for Security Research}

Competition artifacts can serve as reusable benchmarks after the event. Vulnerable RTL designs, known bug sets, security properties, test environments, and exploit cases provide common targets for evaluating new analysis methods. Hack@DAC benchmarks, for example, have been reused to evaluate static analysis, information-flow tracking, symbolic and concolic execution, formal verification, and LLM-assisted bug detection \cite{cweat_paper,lasso,cellift,marvel,symbfuzz}. This enables more consistent comparisons across techniques and reduces dependence on small synthetic benchmarks.

\subsubsection{Security-Aware EDA Tool Development}

The diversity of competition bugs provides useful test cases for evaluating security-aware EDA tools. Existing simulation, lint, and formal flows can be assessed against realistic security failures, while new techniques can be evaluated for detection capability, scalability, false-positive rate, and verification effort. Competition results can therefore reveal gaps in current design-automation flows and motivate tighter integration of security analysis into conventional RTL verification.

\subsubsection{Open Challenges and Research Directions}
Several technical challenges remain for scalable SoC security assurance. First, security intent must be extracted and translated into executable verification properties with limited manual effort. Second, many vulnerabilities require cross-IP and hardware-software interaction that are difficult to capture using module-level verification alone. Third, conventional functional coverage does not directly measure whether security-critical assets, privilege transitions, and adversarial scenarios have been adequately explored. Security-oriented coverage metrics are therefore needed.

Future work should also investigate tighter integration of simulation, formal verification, static analysis, and fuzzing so that results from one technique can guide another. LLM-assisted analysis provides another promising direction, but generated bug hypotheses and properties should remain grounded in RTL semantics and validated through EDA tools. Finally, future competition benchmarks should include larger heterogeneous SoCs, AI accelerators, chiplet-based systems, and cross-abstraction vulnerabilities to better reflect emerging hardware-security challenges.

\section{Conclusion}
\label{sec:conclusion}

Hardware hacking competitions provide a practical environment for evaluating SoC security verification methods against realistic design weaknesses. This paper examined multiple analysis strategies, including simulation-based verification, formal property and security verification, RTL lint analysis, CWE-guided bug detection using LLMs, and hybrid fuzzing. The findings indicate that these techniques provide complementary capabilities, while reliable vulnerability discovery requires explicit security intent, systematic failure analysis, and independent validation of candidate bugs.
Beyond the competition setting, the released designs and known vulnerabilities provide useful benchmarks for evaluating security-aware EDA tools and emerging automated analysis techniques. The lessons from these competitions can therefore help improve pre-silicon vulnerability detection, cross-layer security verification, and reproducible assessment of future SoC security methodologies.

\vspace{-1em}
\section*{Acknowledgments}

The authors would like to thank Dr. Atri Chatterjee, Dr. Rajat Sadhukhan, and Dr. Arnab Bag for their participation and contributions across different competition editions, and Dr. Debdeep Mukhopadhyay for his valuable insights and guidance. The authors also acknowledge the organizers from Texas A\&M University, TU Darmstadt, Intel Corporation, and Synopsys for successfully organizing the competitions over the years and for providing a valuable platform for advancing hardware security research.


\vspace{-1em}
\bibliographystyle{IEEEtran}
\bibliography{IEEEabrv, references}


\end{document}